\documentclass[aps,prl,twocolumn,showpacs,amsmath,amssymb,superscriptaddress,longbibliography]{revtex4-2}
\usepackage{xcolor}
\definecolor{blues}{RGB}{0,108,169}
\definecolor{blaa}{RGB}{153,153,255}
\definecolor{blaa}{RGB}{0,0,125}
\definecolor{filtered}{RGB}{153,0,51}
\definecolor{raw}{RGB}{255,177,100}
\definecolor{filtered}{RGB}{175,0,0}
\definecolor{groen}{RGB}{0,150,0}
\definecolor{lil}{RGB}{160,33,160}
\definecolor{orang}{RGB}{255,60,0}
\usepackage{soul}
\usepackage[normalem]{ulem}

\usepackage{amssymb,amsmath}
\usepackage{siunitx}
\usepackage[colorlinks=true,urlcolor=BrickRed,citecolor=blues,linkcolor=ForestGreen]{hyperref}
\usepackage[resetlabels,labeled]{multibib}
\newcites{S}{References}
\usepackage{upgreek}

\usepackage{graphicx}
\usepackage{epstopdf}
\usepackage{float}
\usepackage[dvipsnames]{xcolor}

\newcommand{\ket}[1]{|#1\rangle}

\newcommand{\eref}[1]{Eq.~(\ref{#1})}

\newcommand{\fref}[1]{Fig.~\ref{#1}}

\newcommand{\Fref}[1]{Figure~\ref{#1}}

\usepackage{orcidlink}
\usepackage{nicematrix}
\NiceMatrixOptions{
	code-for-first-row = \color{black!30!ForestGreen} ,
	code-for-last-row = \color{black!30!ForestGreen} ,
	code-for-first-col = \color{black!30!ForestGreen} ,
	code-for-last-col = \color{black!30!ForestGreen}
}

\makeatletter
\newcommand{\ORCID}[1]{{\orcidlink{#1}}}
\makeatother

\DeclareFontFamily{U}{mathb}{\hyphenchar\font45}
\DeclareFontShape{U}{mathb}{m}{n}{
	<5> <6> <7> <8> <9> <10> gen * mathb
	<10.95> mathb10 <12> <14.4> <17.28> <20.74> <24.88> mathb12
}{}
\DeclareSymbolFont{mathb}{U}{mathb}{m}{n}
\DeclareFontSubstitution{U}{mathb}{m}{n}
\DeclareFontFamily{U}{mathx}{\hyphenchar\font45}
\DeclareFontShape{U}{mathx}{m}{n}{
	<5> <6> <7> <8> <9> <10>
	<10.95> <12> <14.4> <17.28> <20.74> <24.88>
	mathx10
}{}
\DeclareSymbolFont{mathx}{U}{mathx}{m}{n}
\DeclareFontSubstitution{U}{mathx}{m}{n}
\DeclareMathDelimiter{\thickvert}{0}{mathb}{"7E}{mathx}{"1F}

\begin{document}
\title{Observing Autler--Townes splitting of an rf-dressed Feshbach resonance}

\author{Samyajit Gayen\ORCID{0009-0008-8228-2860}}
\affiliation{
	Department of Physics, QSO-Quantum Science Otago, and Dodd-Walls Centre for Photonic and Quantum Technologies,
	University of Otago, Dunedin 9016, New Zealand
}%
\author{Matthew Chilcott\ORCID{0000-0002-1664-6477}}
\affiliation{
	Department of Physics, QSO-Quantum Science Otago, and Dodd-Walls Centre for Photonic and Quantum Technologies,
	University of Otago, Dunedin 9016, New Zealand
}%
\author{Chongwu Xie\ORCID{0009-0005-4601-930X}}
\affiliation{
	Department of Physics, QSO-Quantum Science Otago, and Dodd-Walls Centre for Photonic and Quantum Technologies,
	University of Otago, Dunedin 9016, New Zealand
}%

\author{Niels Kj{\ae}rgaard\ORCID{0000-0002-7830-9468}}%
\email{niels.kjaergaard@otago.ac.nz}
\affiliation{
	Department of Physics, QSO-Quantum Science Otago, and Dodd-Walls Centre for Photonic and Quantum Technologies,
	University of Otago, Dunedin 9016, New Zealand
}%
\date{\today}

\begin{abstract}
We experimentally implement the scheme introduced by Ding, D’Incao, and Greene [\href {https://doi.org/10.1103/physreva.95.022709}{Phys. Rev. A {\bf 95}, 022709
  (2017)}] for controlling the scattering length near a magnetic Feshbach resonance using an rf dressing field.  Specifically, we observe Autler–Townes splitting of the 1007 G Feshbach resonance in dipole-trapped $^{87}\mathrm{Rb}$ by coupling the associated quasibound molecular state to a deeper-lying bound state with a different spin projection. At a fixed magnetic field away from the bare resonance, we further demonstrate rf switching of the collisional response. Our experimental observable is atom loss, which is consistent with enhanced three-body recombination arising from an increase in the magnitude of the real part of the scattering length. Our observation of Autler--Townes splitting establishes the feasibility of generating the strong rf magnetic fields required for sufficient molecular dressing and provides an important basis for further experimental tests of the predicted elastic nature of the rf-controlled interactions.
   \end{abstract}

\maketitle
Ultracold atomic gases provide an exceptionally clean and controllable testbed for exploring quantum many-body physics \cite{Bloch2008}. At the extremely low collision energies characteristic of these systems, interactions between atoms can, to a large extent, be characterized by a single parameter: the $s$-wave scattering length, $a_s$. The scattering length enters prominently into the effective interaction strength and therefore governs many mean-field properties of the gas, including its chemical potential, collective excitations, and stability. A particularly powerful feature of ultracold gases is that the scattering length is not necessarily fixed: through the use of a magnetic Feshbach resonance, an applied magnetic field can tune a molecular bound state relative to the scattering threshold, allowing $a_s$ to be varied over a very wide range and even changed in sign \cite{Tiesinga1993,Inouye1998,RevModPhys.82.1225}.

The ability to control the strength and character of interparticle interactions has made ultracold gases a versatile platform for investigating phenomena ranging from weakly-interacting Bose–Einstein condensates to strongly correlated quantum matter.
\begin{figure}
    \centering
\includegraphics[width=\linewidth]{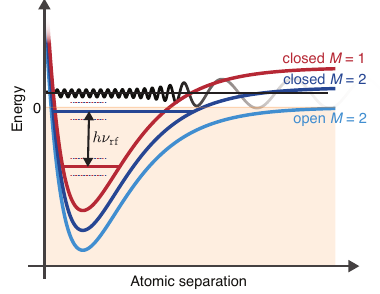}
    \caption{Controlling atomic scattering length with an rf field. At long range, colliding atomic pairs (wave function in black) belong asymptotically to the open $M=2$ channel with each atom in $\ket{1,1}$. At short range, they experience the effect of a field-tunable near-threshold bound state (horizontal dark blue line) that resides over closed channels with $M=2$ and gives rise to a Feshbach resonance as it is tuned across threshold.
    Below the $M=2$ bound state is a bound state belonging to closed channels with $M=1$ (horizontal red line) to which coupling can be instigated with a resonant rf field. The horizontal dashed lines illustrate the dressed states resulting from resonantly coupling the two bound states. The splitting of the Feshbach state into an Autler--Townes doublet affects scattering and in particular $a_s$.   
    }
\label{fig:molecular_potentials}
\end{figure}
However, the speed at which the interaction strength can be varied using a magnetic Feshbach resonance is ultimately limited by the response of the coils that establish the required field. Due to inductance, the current flowing in a coil cannot be changed instantaneously and this becomes particularly restrictive when large magnetic fields are required as is often the case for magnetic Feshbach resonances. A number of strategies have been presented to obtain fast control over $a_s$ \cite{Fedichev1996,Theis2004,Zhang2009,Bauer2009,Tscherbul2010,Hanna2010,Papoular2010,Smith2015,Nicholson2015,Jagannathan2016,Owens2016,Ding2017,Bentine2020,Wang2025}, for example motivated by the ability to perform quenching experiments \cite{Journeaux2026} or to Floquet-engineer interatomic interactions \cite{Sykes2017,Dauer2025}.

One route to overcoming the bandwidth limitation encountered when using dc magnetic fields for tuning is to control the scattering 
length using an additional rf magnetic field while keeping the static 
bias field fixed. At least two complementary experimental approaches have 
demonstrated the potential of rf-controlled atomic interactions. Kaufman \textit{et al.} \cite{Kaufman2009} used rf radiation to couple a manifold of 
several closely spaced molecular bound states associated with Feshbach  resonances near 9~G in $^{87}$Rb. The magnetic component of the rf field was perpendicular to that of the bias field and the resulting multistate dressing produced a series of avoided crossings and pronounced modifications of the scattering 
properties. These low-field resonances, however, are accompanied by 
substantial inelastic spin-relaxation \cite{Sawyer2017}, making them less favorable for 
applications requiring coherent and low-loss control of the interaction 
strength. More recently, Guthmann \textit{et al.}~\cite{Guthmann2025} 
demonstrated a conceptually different approach in $^{6}$Li, using strong 
periodic modulation along the direction of the magnetic bias field to ``Floquet-engineer'' a molecular dimer state. New scattering resonances appear when these dressed molecular levels cross 
the atomic threshold, with their positions and properties controlled through 
both the modulation frequency and amplitude. Inducing and manipulating Feshbach resonances through rf fields offers considerable scope as was recently underscored by the realization of a Friedrich--Wintgen bound state in the continuum \cite{Guthmann2026}.

A particularly simple implementation of rf control was proposed by Ding 
\textit{et al.}~\cite{Ding2017}, where a low-loss magnetic 
Feshbach state near threshold was rf-coupled to an isolated lower-lying molecular bound state (see \fref{fig:molecular_potentials}). The resonant rf dressing of the system is predicted to split and shift the Feshbach resonance experienced by atom pairs in the entrance channel. The rf field hence provides direct control of the scattering length and since parameters such as frequency and amplitude can be varied electronically 
on time scales far shorter than those associated with a large bias-field, the proposal provides a route to versatile control of the interaction strength. 

In this Letter, we experimentally realize 
the proposal of Ref.~\citenum{Ding2017} by demonstrating an Autler--Townes-like splitting \cite{Autler1955,CohenTannoudji1996} of an elastic, closed-channel-dominated Feshbach resonance, in effect controlling atomic interactions through an rf field. We specifically address the widely-used $s$-wave Feshbach resonance of $\rm^{87}Rb$ near $B=1007~\mathrm{G}$~ \cite{PhysRevLett.89.283202}, associated with the 
scattering channel connecting asymptotically to free pairs of atoms in the ground state. At $B=0$, this corresponds to the atomic $\ket{F=1,m_F=1}$ hyperfine state. This open channel---illustrated in \fref{fig:molecular_potentials} by a diabatic potential in light blue and a scattering wavefunction in black---has $M=m_{F_1}+m_{F_2}=2$. The bound state giving rise to the Feshbach resonance is represented schematically in dark blue and has components from multiple closed channels \footnote{Ref.~\citenum{Ding2017} specifically attributes the Feshbach resonance to a bound state in the channel $(F_1m_{F_1},F_2m_{F_2})=(20,22)$, where, in reality, the bound state also has components from $(11,21)$, $(10,22)$, and $(21,21)$}, all with $M=2$---a conserved quantity for collisions in a magnetic field \cite{RevModPhys.82.1225}. It is the coupling between the open channel and closed $M=2$ channels at short range which gives rise to the magnetically tunable Feshbach resonance.  \Fref{fig:molecular_potentials} also presents a bound state residing across closed channels with $M=1$ (red line), which, in itself, does not couple to $M=2$ to affect the Feshbach interaction. We shall demonstrate that by applying an rf field of the appropriate polarization and sufficient amplitude to resonantly drive a bound-bound transition between $M=2$ and $M=1$, we can control the collisional interaction via the resulting Autler--Townes splitting of the bare bound states.

\begin{figure}
    \centering
\includegraphics[width=\linewidth]{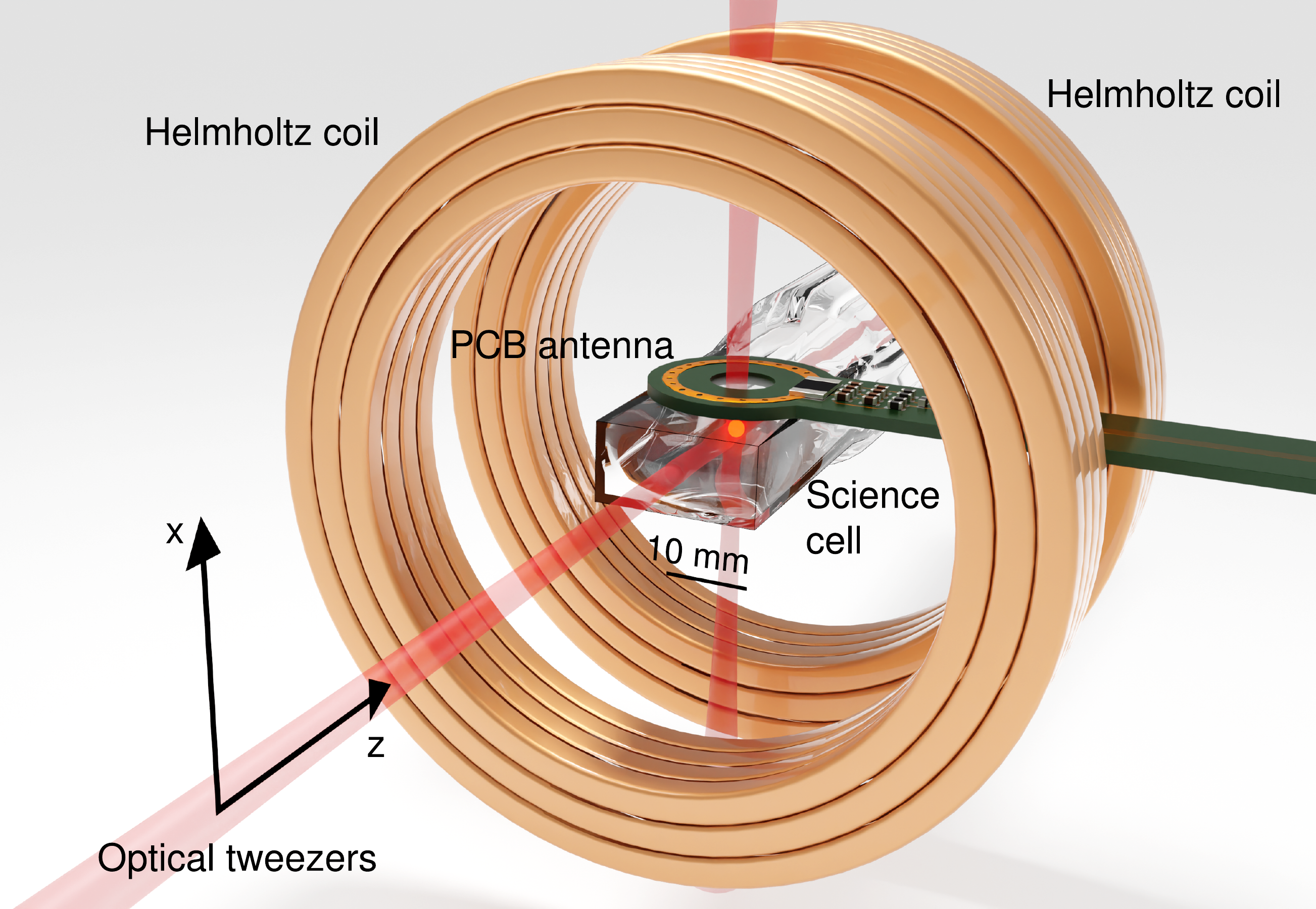}
    \caption{Annotated scale drawing of experimental setup.  At the location of the atoms, the dc bias magnetic field from the Helmholtz coils is along the $z$-axis (horizontal) and the rf magnetic field from the PCB loop antenna is in the direction of the $x$-axis (vertical).  \label{fig:setup}}
    \label{fig:placeholder}
\end{figure}

As a benchmark, we locate and characterize the bare Feshbach resonance in the absence of rf dressing using loss spectroscopy. \Fref{fig:setup} shows our experimental setup, where a $\sim \SI{2}{\micro\kelvin}$ sample of up to $4\times10^{6}$ $\rm^{87}Rb$ atoms in the $\ket{1,1}$ hyperfine state is confined by a crossed dipole trap inside an evacuated glass cell (see End Matter). Since $\ket{1,1}$ constitutes the absolute atomic Zeeman ground state, pairs of atoms do not suffer from two-body inelastic loss and are described by an entirely real scattering length \cite{Hutson2007}. An increase in $|a_s|$ however leads to a knock-on effect in the form of enhanced three-body loss and is hence anticorrelated to the number of atoms remaining in the trap after some hold time $t_{\rm hold}$. We generate a dc magnetic bias field $B_{\rm bias}$ using a pair of water-cooled coils in a Helmholtz configuration with currents actively regulated at the sub-ppm level \cite{Thomas2020}. The direction of the bias field is along the horizontal trapping beam (see \fref{fig:setup}), which we take as the quantization axis for our system ($z$-axis).
  \begin{figure}
\includegraphics[width=0.95\linewidth]{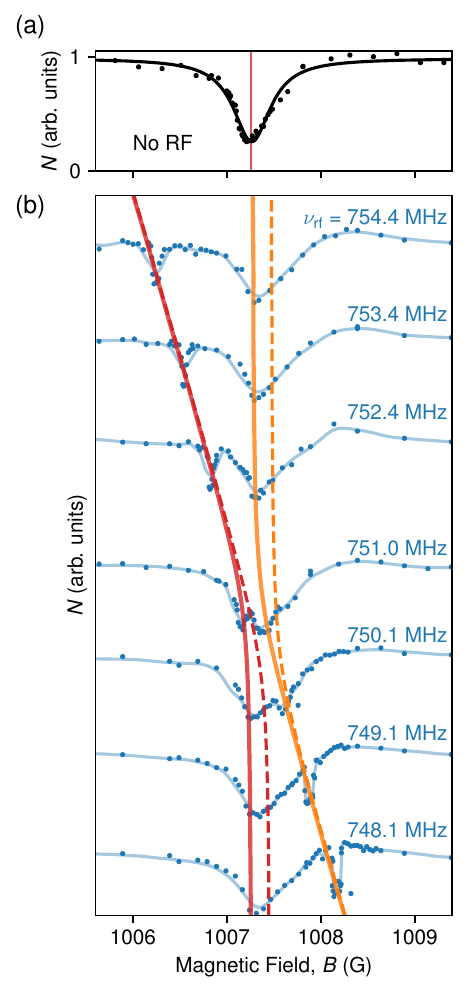}
    \caption{Radio-frequency-dressed splitting of a Feshbach resonance. (a) Bare resonance measured in loss spectroscopy with data fitted using the finite-temperature model of Ref.~\citenum{PhysRevA.106.023303}. The vertical red line marks the zero-energy Feshbach resonance position. (b) Loss spectra acquired in the presence of an rf field and vertically offset according to the applied rf frequency. Blue circles denote the measurements, and the light-blue curves are guides to the eye. The superimposed red and orange curves show the calculated magnetic-field positions of the two rf-dressed features for a simulated rf magnetic-field amplitude of $B_{\mathrm{rf}}=1.14~\mathrm{G}$. Solid curves denote maxima of the calculated cross section, while dashed curves denote the corresponding minima. Application of the rf field splits the bare Feshbach resonance, with the observed branches following the calculated rf-dressed resonance structure.
    \label{fig:loss}}
\end{figure}
 While holding the atomic cloud in the optical
  dipole trap, the magnetic field is brought from near-zero to a selected value in the vicinity of the Feshbach resonance in 65~ms using a minimum-jerk ramp \cite{Flash1985}. After a 32~ms hold time, we simultaneously turn off the  Helmholtz field and the optical trapping light. The number of remaining atoms $N$ is determined using absorption imaging following a 20~ms time-of-flight period. \Fref{fig:loss}(a) shows $N$ versus the holding field and the clear manifestation of the Feshbach resonance loss feature at $\sim1007.3$~G. 
\begin{figure}
    \centering
\includegraphics[width=\linewidth]{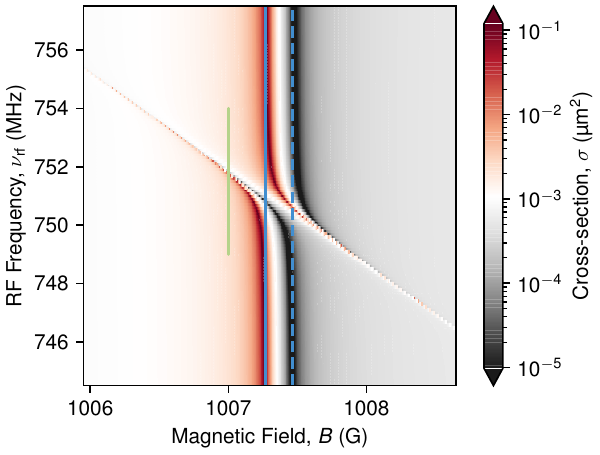}
    \caption{Calculated scattering cross section as a function of rf frequency $\nu_{\mathrm{rf}}$ and applied magnetic field $B$, evaluated at a collision energy of $k_B\times1\upmu$K and a dressing field amplitude $B_{\rm rf}=1.14$~G. The maxima (minima) of the two branches of the rf-dressed Feshbach resonance appear as red (black) features. The solid (dashed) blue line marks  the position of the resonance maximum (minimum) in the absence of rf coupling. The solid green line indicates the iso-$B$ slice explored experimentally in \fref{fig:rf_loss_B1007G}c.}
\label{fig:CrossSections_with_Bare}
\end{figure}

We next turn to resonantly coupling the $M=2$ Feshbach state to
a deeper-lying bound state with $M=1$. Using the potentials of Ref.~\onlinecite{PhysRevA.82.052514},  we calculate the energy variation as a function of magnetic field for the bound states of $\rm^{87}Rb$ (see End Matter). At the magnetic field for which the bound $M=2$ state crosses the $\ket{1,1}+\ket{1,1}$ threshold (i.e., $\sim 1007.3$~G), we find an $M=1$ bound state residing at an energy $h\times 751$~ MHz below. This is nominally the same $M=1$ bound state considered in the proposal of Ref.~\citenum{Ding2017} \footnote{Ref.~\citenum{Ding2017} states a transition frequency $\nu=745$~ MHz, slightly lower than what we find, which we attribute to using different interaction potentials for the close-coupled calculations. We also note that it designates the molecular state as $(2-1,22)$, where, in fact, components from $(1-1,22)$,$(10,11)$,$(11,20)$,$(21,20)$, and $(11,10)$ will also contribute.}. Our survey also shows additional bound $M=1$ states at $\{+585, +43, -31, -212\}$~MHz with respect to the $\ket{1,1}+\ket{1,1}$ threshold, but these all have a significantly smaller magnetic dipole coupling to the $M=2$ state.

To achieve the required change of one unit of angular momentum associated with an $M=2\leftrightarrow M=1$ magnetic dipole transition, we apply an rf field with its magnetic component pointing along the vertical $x$-direction, i.e., perpendicularly to our chosen quantization axis \cite{Vanier1989}.
The proposal of Ref.~\citenum{Ding2017} considers a dressing field of amplitude $B_{\rm rf}=5$~G for which an Autler--Townes splitting of $\sim0.5$~G (inferred from Fig. 2 of Ref.~\citenum{Ding2017}) is predicted. Outside the field of atom chips, it is technologically non-trivial to achieve an rf field of such magnitude at the heart of an ultracold atom machine and this might explain why the scheme has been awaiting an experimental realization for more than a decade---rf fields achieved in setups with ultracold atoms typically fall an order of magnitude below this number \cite{Cosco2023,Scazza2025,Jain2026}. We employ a custom-built printed-circuit-board (PCB) loop antenna, inspired by designs developed for RFID applications in the $900$--$930$~MHz band. By suitable rescaling of loop dimensions and modification of matching network components, we achieve operation over the band $\nu_{\rm rf}=745$--$763~\mathrm{MHz}$ (see End Matter) capable of handling an input power of 100~W for at least 50~ms. The PCB design allows for installation of the loop flush against the science cell at a $\sim 6$~mm distance from the atoms (see \fref{fig:setup}). For an input power of $100~\mathrm{W}$, we achieve an amplitude $B_{\mathrm{rf}}=1.14(11)~\mathrm{G}$ for the magnetic field $B_{\mathrm{rf}}\cos(2\pi\nu_{\rm rf}t)\hat{\bf x}$, at the location of the atoms as estimated from hyperfine Rabi spectroscopy (see End Matter). We use the measured value $B_{\mathrm{rf}}=1.14~\mathrm{G}$ to predict the variation in elastic cross section $\sigma$ for a parameter space spanned by $B_{\rm bias}$ and the dressing field frequency $\nu_{\rm rf}$ (see \fref{fig:CrossSections_with_Bare}). In particular, our calculations at $B_{\mathrm{rf}}=1.14~\mathrm{G}$ predict a minimum field-dressed splitting of the bare Feshbach resonance $\Delta_{\rm AT}=0.233
$~G  for $\nu_{\rm rf}=751$~MHz. 
 
\begin{figure}[t!] 
\centering \includegraphics[width=\columnwidth]{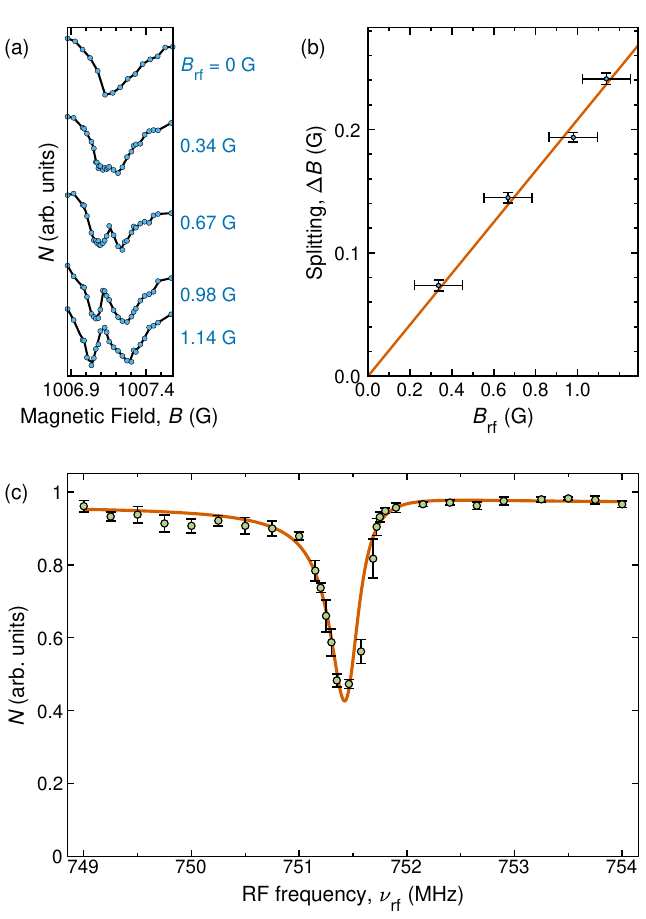} 
\caption{(a) Loss spectrum development as the amplitude of a resonant dressing field ($\nu_{\rm rf}=\nu_0$) increases from $B_{\rm rf}=0$~G (no rf field applied) to $B_{\rm rf}=1.14$~G.
(b) Resonance-position separation $\Delta B$ between the two rf-induced loss features as a function of $B_{\rm rf}$. The feature positions $B_{1}$ and $B_{2}$ were determined from independent local Lorentzian fits to the two loss minima, and the separation was calculated as $\Delta B=|B_{2}-B_{1}|$. Vertical error bars are the propagated standard errors of the fitted centers. The horizontal error bars are the standard deviation of $B_{\rm rf}$ between different time windows of the rf pulse, measured at maximum drive and used as an upper bound at all powers. The solid line is a linear fit to the measured separation.
(c)  rf-induced loss spectrum measured at a magnetic field of $B = 1007.0~\mathrm{G}$, away from the bare Feshbach resonance. The remaining atom fraction is plotted as a function of the applied rf frequency. The solid curve is a fit of a Beutler--Fano profile to the loss feature, yielding an rf-induced resonance frequency of $\nu_{\mathrm{res}} = 751.447(8)~\mathrm{MHz}$.} \label{fig:rf_loss_B1007G}
\end{figure}

  To observe the predicted induced Autler--Townes splitting of the Feshbach resonance in \fref{fig:loss}(a) experimentally, we use our PCB antenna to rf dress the atoms. We add to the loss spectroscopy sequence the application of a fixed-frequency dressing field 2~ms into the holding stage. The dressing field is switched off simultaneously with the dipole trap and magnetic bias field at the end of holding. \Fref{fig:loss}(b) presents the loss curves acquired for seven values of $\nu_{\rm rf}$, distributed about $751$~MHz---the calculated frequency for which an rf photon matches the energy gap from the $M=1$ bound state to the $M=2$ Feshbach state at threshold (i.e., $B=1007.3$~G). In all cases the rf dressing gives rise to a clear splitting of the Feshbach resonance. We overlay the experimental data with the predicted splittings from our calculations (cf. \fref{fig:CrossSections_with_Bare}) and note a good correspondence to the observed variation in splitting with $\nu_{\rm rf}$.
 Modeling the splitting to have the generalized Rabi form
$\Delta B \propto \sqrt{(\Omega_0)^2 + [2\pi(\nu_{\rm rf} - \nu_0)]^2}$, we obtain $\nu_0=750.99(4)$~MHz, in excellent agreement with our calculations.

At resonance ($\nu_{\rm rf} = \nu_0$), the splitting $\Delta B$ is proportional to the bound-bound Rabi frequency $\Omega_0$, which in turn is expected to be proportional to the rf magnetic field $B_\text{rf}$ (see End Matter).
\Fref{fig:rf_loss_B1007G}(a) shows the development in experimental loss spectra at
$\nu_{\rm rf}=750.99$~MHz as the rf magnetic field is increased, while \fref{fig:rf_loss_B1007G}(b) presents the corresponding inferred splittings in minima. The splitting is well described by $\Delta B=\alpha B_{\rm rf}$ with a constant of proportionality $\alpha=0.208(15)$, in agreement with the calculated value $\alpha_{\rm calc}=0.205$ from a two-level model (see End Matter).
The maximum resonant Autler–Townes splitting achievable in our current setup is $\sim$ 0.24~G, comparable to the thermally broadened width of the bare Feshbach loss feature, $\sim$0.25~G. The overlap between the resonantly dressed branches and the bare resonance restricts our ability to switch between strongly and weakly interacting conditions at $B=1007.3$~G. For the purpose of demonstrating a switching action, we instead operate at a fixed bias field $B=1007.0$~G, where the bare resonance produces little additional loss and our calculations (\fref{fig:CrossSections_with_Bare}) predict an rf-induced resonance near $\nu_{\rm rf}=751.45$~MHz.
Conducting loss spectroscopy at $B=1007.0~\mathrm{G}$ and varying the frequency of the dressing field along the green line in \fref{fig:CrossSections_with_Bare}, we indeed observe a clear loss feature at $\nu_{\mathrm{res}} = 751.447(8)~\mathrm{MHz}$ [see \fref{fig:rf_loss_B1007G}(c)]. This demonstrates the ability to turn on and off the collisional interaction through a field of $\nu_{\rm rf} = \nu_{\rm res}$.

In conclusion, we have observed Autler--Townes splitting of a Feshbach resonance by dressing the rf transition from the Feshbach state to a lower-lying molecular state. This verifies predictions from the proposal by Ding, D'Incao, and Greene, where this particular system is used as a case study for rf control of the scattering length via dressing of a bound-bound transition.

\begin{figure}[b!] \includegraphics[width=\linewidth]{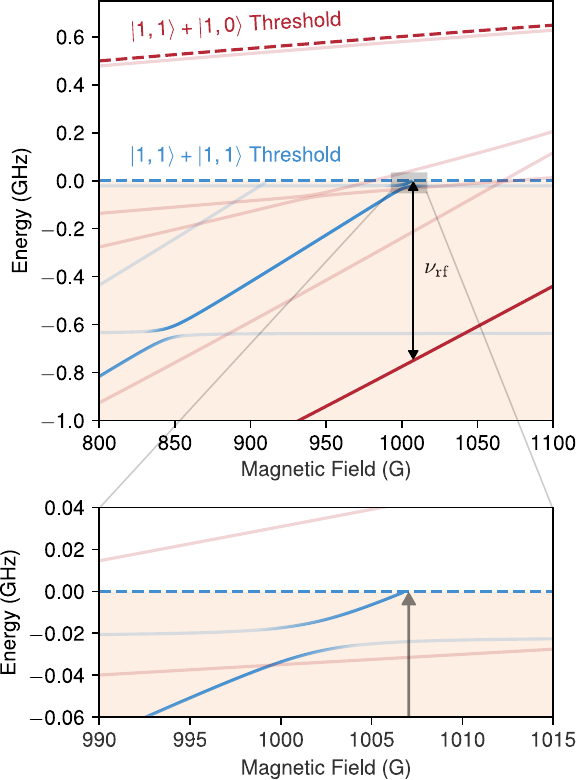}
    \caption{Calculated $\rm^{87}Rb_2$  bound states. An rf field of frequency $\nu_{\rm rf}\sim 750$~MHz couples the $M=2$ Feshbach state that crosses the $\ket{1,1}+\ket{1,1}$ threshold (blue dashed line) for a $\sim1007$~G magnetic field and a deeper-lying bound state belonging to the manifold of $M=1$ channels. Out of the $M=1$ manifold the $\ket{1,1}+\ket{1,0}$ channel has the lowest threshold (dashed red line). 
     \label{fig:bound_states}}
\end{figure}

We clearly demonstrate control over the collisional interactions, but to validate that our observation of atomic loss results from an increase in the real part of the scattering length would require further experiments. While there is no inelastic two-body loss path for bare $\ket{1,1}$ atoms, in principle, our dressing rf field not only couples $\ket{1,1}$+$\ket{1,1}$ to the discrete bound $M=1$ state via the $M=2$ Feshbach state, but also to the continuum above the $\ket{1,1}$+$\ket{1,0}$ threshold (see \fref{fig:bound_states} in End Matter) \footnote{For a field  $B=1007.3$~G, the $\ket{1,1}+\ket{1,0}$ threshold is located $\sim 605$~MHz above $\ket{1,1}+\ket{1,1}$, making the transition energetically allowed for a 751~MHz photon.}.

According to Ref.~\onlinecite{Ding2017}, rf dissociation into inelastic channels is negligible for this system, and our own coupled-channels calculations support this conclusion. However, this point warrants further experimental investigation, and in the future we plan to conduct Ramsey spectroscopy \cite{Bennett2017} to ascertain the predicted elastic nature of the enhanced interactions. We will also explore the scope for increasing the amplitude of the dressing field in order to create a resonant Autler--Townes splitting exceeding the width of the bare Feshbach resonance. The required factor of two in $B_{\rm rf}$ might straightforwardly be obtained with a synchronously driven twin antenna located below the science cell \footnote{For a $B_{\rm rf}=2.2$~G amplitude both our calculations and our measured tuning rate $\alpha$ predict a splitting of $\sim 0.5$~G. This mutual agreement does not extend to the prediction of Ref.~\onlinecite{Ding2017}, where a $\sim 0.5$~G splitting is obtained for $B_{\rm rf}=5$~G.}. This should resolve the resonant Autler--Townes doublet beyond the thermal width of the Feshbach resonance and provide a more favorable switching and modulation scenario than in the off-resonant proof-of-concept demonstration of this Letter.

\textit{Acknowledgments---}This work was supported by the Marsden Fund of New
Zealand (Contracts No. UOO1923 and No. UOO2421). The rf loop antenna was inspired by a design by Veljko Ignjatovic. We thank Yijue Ding for input on calculations, and J\'{e}r\^{o}me Beugnon and Aur\'{e}lien Perrin for sharing their insights on the prospects for using free-bound transitions for Feshbach resonance control \cite{Maury2023,Ballu2025}.
\bibliography{references}

\begin{thebibliography}{46}%
\makeatletter
\providecommand \@ifxundefined [1]{%
 \@ifx{#1\undefined}
}%
\providecommand \@ifnum [1]{%
 \ifnum #1\expandafter \@firstoftwo
 \else \expandafter \@secondoftwo
 \fi
}%
\providecommand \@ifx [1]{%
 \ifx #1\expandafter \@firstoftwo
 \else \expandafter \@secondoftwo
 \fi
}%
\providecommand \natexlab [1]{#1}%
\providecommand \enquote  [1]{``#1''}%
\providecommand \bibnamefont  [1]{#1}%
\providecommand \bibfnamefont [1]{#1}%
\providecommand \citenamefont [1]{#1}%
\providecommand \href@noop [0]{\@secondoftwo}%
\providecommand \href [0]{\begingroup \@sanitize@url \@href}%
\providecommand \@href[1]{\@@startlink{#1}\@@href}%
\providecommand \@@href[1]{\endgroup#1\@@endlink}%
\providecommand \@sanitize@url [0]{\catcode `\\12\catcode `\$12\catcode
  `\&12\catcode `\#12\catcode `\^12\catcode `\_12\catcode `\%12\relax}%
\providecommand \@@startlink[1]{}%
\providecommand \@@endlink[0]{}%
\providecommand \url  [0]{\begingroup\@sanitize@url \@url }%
\providecommand \@url [1]{\endgroup\@href {#1}{\urlprefix }}%
\providecommand \urlprefix  [0]{URL }%
\providecommand \Eprint [0]{\href }%
\providecommand \doibase [0]{https://doi.org/}%
\providecommand \selectlanguage [0]{\@gobble}%
\providecommand \bibinfo  [0]{\@secondoftwo}%
\providecommand \bibfield  [0]{\@secondoftwo}%
\providecommand \translation [1]{[#1]}%
\providecommand \BibitemOpen [0]{}%
\providecommand \bibitemStop [0]{}%
\providecommand \bibitemNoStop [0]{.\EOS\space}%
\providecommand \EOS [0]{\spacefactor3000\relax}%
\providecommand \BibitemShut  [1]{\csname bibitem#1\endcsname}%
\let\auto@bib@innerbib\@empty
\bibitem [{\citenamefont {Bloch}\ \emph {et~al.}(2008)\citenamefont {Bloch},
  \citenamefont {Dalibard},\ and\ \citenamefont {Zwerger}}]{Bloch2008}%
  \BibitemOpen
  \bibfield  {author} {\bibinfo {author} {\bibfnamefont {I.}~\bibnamefont
  {Bloch}}, \bibinfo {author} {\bibfnamefont {J.}~\bibnamefont {Dalibard}},\
  and\ \bibinfo {author} {\bibfnamefont {W.}~\bibnamefont {Zwerger}},\
  }\bibfield  {title} {\bibinfo {title} {Many-body physics with ultracold
  gases},\ }\href {https://doi.org/10.1103/revmodphys.80.885} {\bibfield
  {journal} {\bibinfo  {journal} {Rev. Mod. Phys.}\ }\textbf {\bibinfo {volume}
  {80}},\ \bibinfo {pages} {885} (\bibinfo {year} {2008})}\BibitemShut
  {NoStop}%
\bibitem [{\citenamefont {Tiesinga}\ \emph {et~al.}(1993)\citenamefont
  {Tiesinga}, \citenamefont {Verhaar},\ and\ \citenamefont
  {Stoof}}]{Tiesinga1993}%
  \BibitemOpen
  \bibfield  {author} {\bibinfo {author} {\bibfnamefont {E.}~\bibnamefont
  {Tiesinga}}, \bibinfo {author} {\bibfnamefont {B.~J.}\ \bibnamefont
  {Verhaar}},\ and\ \bibinfo {author} {\bibfnamefont {H.~T.~C.}\ \bibnamefont
  {Stoof}},\ }\bibfield  {title} {\bibinfo {title} {Threshold and resonance
  phenomena in ultracold ground-state collisions},\ }\href
  {https://doi.org/10.1103/physreva.47.4114} {\bibfield  {journal} {\bibinfo
  {journal} {Phys. Rev. A}\ }\textbf {\bibinfo {volume} {47}},\ \bibinfo
  {pages} {4114} (\bibinfo {year} {1993})}\BibitemShut {NoStop}%
\bibitem [{\citenamefont {Inouye}\ \emph {et~al.}(1998)\citenamefont {Inouye},
  \citenamefont {Andrews}, \citenamefont {Stenger}, \citenamefont {Miesner},
  \citenamefont {Stamper-Kurn},\ and\ \citenamefont {Ketterle}}]{Inouye1998}%
  \BibitemOpen
  \bibfield  {author} {\bibinfo {author} {\bibfnamefont {S.}~\bibnamefont
  {Inouye}}, \bibinfo {author} {\bibfnamefont {M.~R.}\ \bibnamefont {Andrews}},
  \bibinfo {author} {\bibfnamefont {J.}~\bibnamefont {Stenger}}, \bibinfo
  {author} {\bibfnamefont {H.-J.}\ \bibnamefont {Miesner}}, \bibinfo {author}
  {\bibfnamefont {D.~M.}\ \bibnamefont {Stamper-Kurn}},\ and\ \bibinfo {author}
  {\bibfnamefont {W.}~\bibnamefont {Ketterle}},\ }\bibfield  {title} {\bibinfo
  {title} {Observation of {Feshbach} resonances in a
  {Bose}{\textendash}{Einstein} condensate},\ }\href
  {https://doi.org/10.1038/32354} {\bibfield  {journal} {\bibinfo  {journal}
  {Nature}\ }\textbf {\bibinfo {volume} {392}},\ \bibinfo {pages} {151}
  (\bibinfo {year} {1998})}\BibitemShut {NoStop}%
\bibitem [{\citenamefont {Chin}\ \emph {et~al.}(2010)\citenamefont {Chin},
  \citenamefont {Grimm}, \citenamefont {Julienne},\ and\ \citenamefont
  {Tiesinga}}]{RevModPhys.82.1225}%
  \BibitemOpen
  \bibfield  {author} {\bibinfo {author} {\bibfnamefont {C.}~\bibnamefont
  {Chin}}, \bibinfo {author} {\bibfnamefont {R.}~\bibnamefont {Grimm}},
  \bibinfo {author} {\bibfnamefont {P.}~\bibnamefont {Julienne}},\ and\
  \bibinfo {author} {\bibfnamefont {E.}~\bibnamefont {Tiesinga}},\ }\bibfield
  {title} {\bibinfo {title} {Feshbach resonances in ultracold gases},\ }\href
  {https://doi.org/10.1103/RevModPhys.80.885} {\bibfield  {journal} {\bibinfo
  {journal} {Rev. Mod. Phys.}\ }\textbf {\bibinfo {volume} {82}},\ \bibinfo
  {pages} {1225} (\bibinfo {year} {2010})}\BibitemShut {NoStop}%
\bibitem [{\citenamefont {Fedichev}\ \emph {et~al.}(1996)\citenamefont
  {Fedichev}, \citenamefont {Kagan}, \citenamefont {Shlyapnikov},\ and\
  \citenamefont {Walraven}}]{Fedichev1996}%
  \BibitemOpen
  \bibfield  {author} {\bibinfo {author} {\bibfnamefont {P.~O.}\ \bibnamefont
  {Fedichev}}, \bibinfo {author} {\bibfnamefont {Y.}~\bibnamefont {Kagan}},
  \bibinfo {author} {\bibfnamefont {G.~V.}\ \bibnamefont {Shlyapnikov}},\ and\
  \bibinfo {author} {\bibfnamefont {J.~T.~M.}\ \bibnamefont {Walraven}},\
  }\bibfield  {title} {\bibinfo {title} {Influence of nearly resonant light on
  the scattering length in low-temperature atomic gases},\ }\href
  {https://doi.org/10.1103/physrevlett.77.2913} {\bibfield  {journal} {\bibinfo
   {journal} {Phys. Rev. Lett.}\ }\textbf {\bibinfo {volume} {77}},\ \bibinfo
  {pages} {2913} (\bibinfo {year} {1996})}\BibitemShut {NoStop}%
\bibitem [{\citenamefont {Theis}\ \emph {et~al.}(2004)\citenamefont {Theis},
  \citenamefont {Thalhammer}, \citenamefont {Winkler}, \citenamefont {Hellwig},
  \citenamefont {Ruff}, \citenamefont {Grimm},\ and\ \citenamefont
  {Denschlag}}]{Theis2004}%
  \BibitemOpen
  \bibfield  {author} {\bibinfo {author} {\bibfnamefont {M.}~\bibnamefont
  {Theis}}, \bibinfo {author} {\bibfnamefont {G.}~\bibnamefont {Thalhammer}},
  \bibinfo {author} {\bibfnamefont {K.}~\bibnamefont {Winkler}}, \bibinfo
  {author} {\bibfnamefont {M.}~\bibnamefont {Hellwig}}, \bibinfo {author}
  {\bibfnamefont {G.}~\bibnamefont {Ruff}}, \bibinfo {author} {\bibfnamefont
  {R.}~\bibnamefont {Grimm}},\ and\ \bibinfo {author} {\bibfnamefont {J.~H.}\
  \bibnamefont {Denschlag}},\ }\bibfield  {title} {\bibinfo {title} {Tuning the
  scattering length with an optically induced {Feshbach} resonance},\ }\href
  {https://doi.org/10.1103/physrevlett.93.123001} {\bibfield  {journal}
  {\bibinfo  {journal} {Phys. Rev. Lett.}\ }\textbf {\bibinfo {volume} {93}},\
  \bibinfo {pages} {123001} (\bibinfo {year} {2004})}\BibitemShut {NoStop}%
\bibitem [{\citenamefont {Zhang}\ \emph {et~al.}(2009)\citenamefont {Zhang},
  \citenamefont {Naidon},\ and\ \citenamefont {Ueda}}]{Zhang2009}%
  \BibitemOpen
  \bibfield  {author} {\bibinfo {author} {\bibfnamefont {P.}~\bibnamefont
  {Zhang}}, \bibinfo {author} {\bibfnamefont {P.}~\bibnamefont {Naidon}},\ and\
  \bibinfo {author} {\bibfnamefont {M.}~\bibnamefont {Ueda}},\ }\bibfield
  {title} {\bibinfo {title} {Independent control of scattering lengths in
  multicomponent quantum gases},\ }\href
  {https://doi.org/10.1103/physrevlett.103.133202} {\bibfield  {journal}
  {\bibinfo  {journal} {Phys. Rev. Lett.}\ }\textbf {\bibinfo {volume} {103}},\
  \bibinfo {pages} {133202} (\bibinfo {year} {2009})}\BibitemShut {NoStop}%
\bibitem [{\citenamefont {Bauer}\ \emph {et~al.}(2009)\citenamefont {Bauer},
  \citenamefont {Lettner}, \citenamefont {Vo}, \citenamefont {Rempe},\ and\
  \citenamefont {Dürr}}]{Bauer2009}%
  \BibitemOpen
  \bibfield  {author} {\bibinfo {author} {\bibfnamefont {D.~M.}\ \bibnamefont
  {Bauer}}, \bibinfo {author} {\bibfnamefont {M.}~\bibnamefont {Lettner}},
  \bibinfo {author} {\bibfnamefont {C.}~\bibnamefont {Vo}}, \bibinfo {author}
  {\bibfnamefont {G.}~\bibnamefont {Rempe}},\ and\ \bibinfo {author}
  {\bibfnamefont {S.}~\bibnamefont {Dürr}},\ }\bibfield  {title} {\bibinfo
  {title} {Control of a magnetic {Feshbach} resonance with laser light},\
  }\href {https://doi.org/10.1038/nphys1232} {\bibfield  {journal} {\bibinfo
  {journal} {Nat. Phys.}\ }\textbf {\bibinfo {volume} {5}},\ \bibinfo {pages}
  {339} (\bibinfo {year} {2009})}\BibitemShut {NoStop}%
\bibitem [{\citenamefont {Tscherbul}\ \emph {et~al.}(2010)\citenamefont
  {Tscherbul}, \citenamefont {Calarco}, \citenamefont {Lesanovsky},
  \citenamefont {Krems}, \citenamefont {Dalgarno},\ and\ \citenamefont
  {Schmiedmayer}}]{Tscherbul2010}%
  \BibitemOpen
  \bibfield  {author} {\bibinfo {author} {\bibfnamefont {T.~V.}\ \bibnamefont
  {Tscherbul}}, \bibinfo {author} {\bibfnamefont {T.}~\bibnamefont {Calarco}},
  \bibinfo {author} {\bibfnamefont {I.}~\bibnamefont {Lesanovsky}}, \bibinfo
  {author} {\bibfnamefont {R.~V.}\ \bibnamefont {Krems}}, \bibinfo {author}
  {\bibfnamefont {A.}~\bibnamefont {Dalgarno}},\ and\ \bibinfo {author}
  {\bibfnamefont {J.}~\bibnamefont {Schmiedmayer}},\ }\bibfield  {title}
  {\bibinfo {title} {rf-field-induced {Feshbach} resonances},\ }\href
  {https://doi.org/10.1103/physreva.81.050701} {\bibfield  {journal} {\bibinfo
  {journal} {Phys. Rev. A}\ }\textbf {\bibinfo {volume} {81}},\ \bibinfo
  {pages} {050701} (\bibinfo {year} {2010})}\BibitemShut {NoStop}%
\bibitem [{\citenamefont {Hanna}\ \emph {et~al.}(2010)\citenamefont {Hanna},
  \citenamefont {Tiesinga},\ and\ \citenamefont {Julienne}}]{Hanna2010}%
  \BibitemOpen
  \bibfield  {author} {\bibinfo {author} {\bibfnamefont {T.~M.}\ \bibnamefont
  {Hanna}}, \bibinfo {author} {\bibfnamefont {E.}~\bibnamefont {Tiesinga}},\
  and\ \bibinfo {author} {\bibfnamefont {P.~S.}\ \bibnamefont {Julienne}},\
  }\bibfield  {title} {\bibinfo {title} {Creation and manipulation of
  {Feshbach} resonances with radiofrequency radiation},\ }\href
  {https://doi.org/10.1088/1367-2630/12/8/083031} {\bibfield  {journal}
  {\bibinfo  {journal} {New J. Phys.}\ }\textbf {\bibinfo {volume} {12}},\
  \bibinfo {pages} {083031} (\bibinfo {year} {2010})}\BibitemShut {NoStop}%
\bibitem [{\citenamefont {Papoular}\ \emph {et~al.}(2010)\citenamefont
  {Papoular}, \citenamefont {Shlyapnikov},\ and\ \citenamefont
  {Dalibard}}]{Papoular2010}%
  \BibitemOpen
  \bibfield  {author} {\bibinfo {author} {\bibfnamefont {D.~J.}\ \bibnamefont
  {Papoular}}, \bibinfo {author} {\bibfnamefont {G.~V.}\ \bibnamefont
  {Shlyapnikov}},\ and\ \bibinfo {author} {\bibfnamefont {J.}~\bibnamefont
  {Dalibard}},\ }\bibfield  {title} {\bibinfo {title} {Microwave-induced
  {Fano-Feshbach} resonances},\ }\href
  {https://doi.org/10.1103/physreva.81.041603} {\bibfield  {journal} {\bibinfo
  {journal} {Phys. Rev. A}\ }\textbf {\bibinfo {volume} {81}},\ \bibinfo
  {pages} {041603} (\bibinfo {year} {2010})}\BibitemShut {NoStop}%
\bibitem [{\citenamefont {Smith}(2015)}]{Smith2015}%
  \BibitemOpen
  \bibfield  {author} {\bibinfo {author} {\bibfnamefont {D.~H.}\ \bibnamefont
  {Smith}},\ }\bibfield  {title} {\bibinfo {title} {Inducing resonant
  interactions in ultracold atoms with a modulated magnetic field},\ }\href
  {https://doi.org/10.1103/physrevlett.115.193002} {\bibfield  {journal}
  {\bibinfo  {journal} {Phys. Rev. Lett.}\ }\textbf {\bibinfo {volume} {115}},\
  \bibinfo {pages} {193002} (\bibinfo {year} {2015})}\BibitemShut {NoStop}%
\bibitem [{\citenamefont {Nicholson}\ \emph {et~al.}(2015)\citenamefont
  {Nicholson}, \citenamefont {Blatt}, \citenamefont {Bloom}, \citenamefont
  {Williams}, \citenamefont {Thomsen}, \citenamefont {Ye},\ and\ \citenamefont
  {Julienne}}]{Nicholson2015}%
  \BibitemOpen
  \bibfield  {author} {\bibinfo {author} {\bibfnamefont {T.~L.}\ \bibnamefont
  {Nicholson}}, \bibinfo {author} {\bibfnamefont {S.}~\bibnamefont {Blatt}},
  \bibinfo {author} {\bibfnamefont {B.~J.}\ \bibnamefont {Bloom}}, \bibinfo
  {author} {\bibfnamefont {J.~R.}\ \bibnamefont {Williams}}, \bibinfo {author}
  {\bibfnamefont {J.~W.}\ \bibnamefont {Thomsen}}, \bibinfo {author}
  {\bibfnamefont {J.}~\bibnamefont {Ye}},\ and\ \bibinfo {author}
  {\bibfnamefont {P.~S.}\ \bibnamefont {Julienne}},\ }\bibfield  {title}
  {\bibinfo {title} {{Optical Feshbach resonances: Field-dressed theory and
  comparison with experiments}},\ }\href
  {https://doi.org/10.1103/physreva.92.022709} {\bibfield  {journal} {\bibinfo
  {journal} {Phys. Rev. A}\ }\textbf {\bibinfo {volume} {92}},\ \bibinfo
  {pages} {022709} (\bibinfo {year} {2015})}\BibitemShut {NoStop}%
\bibitem [{\citenamefont {Jagannathan}\ \emph {et~al.}(2016)\citenamefont
  {Jagannathan}, \citenamefont {Arunkumar}, \citenamefont {Joseph},\ and\
  \citenamefont {Thomas}}]{Jagannathan2016}%
  \BibitemOpen
  \bibfield  {author} {\bibinfo {author} {\bibfnamefont {A.}~\bibnamefont
  {Jagannathan}}, \bibinfo {author} {\bibfnamefont {N.}~\bibnamefont
  {Arunkumar}}, \bibinfo {author} {\bibfnamefont {J.}~\bibnamefont {Joseph}},\
  and\ \bibinfo {author} {\bibfnamefont {J.}~\bibnamefont {Thomas}},\
  }\bibfield  {title} {\bibinfo {title} {Optical control of magnetic {Feshbach}
  resonances by closed-channel electromagnetically induced transparency},\
  }\href {https://doi.org/10.1103/physrevlett.116.075301} {\bibfield  {journal}
  {\bibinfo  {journal} {Phys. Rev. Lett.}\ }\textbf {\bibinfo {volume} {116}},\
  \bibinfo {pages} {075301} (\bibinfo {year} {2016})}\BibitemShut {NoStop}%
\bibitem [{\citenamefont {Owens}\ \emph {et~al.}(2016)\citenamefont {Owens},
  \citenamefont {Xie},\ and\ \citenamefont {Hutson}}]{Owens2016}%
  \BibitemOpen
  \bibfield  {author} {\bibinfo {author} {\bibfnamefont {D.~J.}\ \bibnamefont
  {Owens}}, \bibinfo {author} {\bibfnamefont {T.}~\bibnamefont {Xie}},\ and\
  \bibinfo {author} {\bibfnamefont {J.~M.}\ \bibnamefont {Hutson}},\ }\bibfield
   {title} {\bibinfo {title} {Creating {Feshbach} resonances for ultracold
  molecule formation with radio-frequency fields},\ }\href
  {https://doi.org/10.1103/physreva.94.023619} {\bibfield  {journal} {\bibinfo
  {journal} {Phys. Rev. A}\ }\textbf {\bibinfo {volume} {94}},\ \bibinfo
  {pages} {023619} (\bibinfo {year} {2016})}\BibitemShut {NoStop}%
\bibitem [{\citenamefont {Ding}\ \emph {et~al.}(2017)\citenamefont {Ding},
  \citenamefont {D’Incao},\ and\ \citenamefont {Greene}}]{Ding2017}%
  \BibitemOpen
  \bibfield  {author} {\bibinfo {author} {\bibfnamefont {Y.}~\bibnamefont
  {Ding}}, \bibinfo {author} {\bibfnamefont {J.~P.}\ \bibnamefont
  {D’Incao}},\ and\ \bibinfo {author} {\bibfnamefont {C.~H.}\ \bibnamefont
  {Greene}},\ }\bibfield  {title} {\bibinfo {title} {Effective control of cold
  collisions with radio-frequency fields},\ }\href
  {https://doi.org/10.1103/physreva.95.022709} {\bibfield  {journal} {\bibinfo
  {journal} {Phys. Rev. A}\ }\textbf {\bibinfo {volume} {95}},\ \bibinfo
  {pages} {022709} (\bibinfo {year} {2017})}\BibitemShut {NoStop}%
\bibitem [{\citenamefont {Bentine}\ \emph {et~al.}(2020)\citenamefont
  {Bentine}, \citenamefont {Barker}, \citenamefont {Luksch}, \citenamefont
  {Sunami}, \citenamefont {Harte}, \citenamefont {Yuen}, \citenamefont {Foot},
  \citenamefont {Owens},\ and\ \citenamefont {Hutson}}]{Bentine2020}%
  \BibitemOpen
  \bibfield  {author} {\bibinfo {author} {\bibfnamefont {E.}~\bibnamefont
  {Bentine}}, \bibinfo {author} {\bibfnamefont {A.~J.}\ \bibnamefont {Barker}},
  \bibinfo {author} {\bibfnamefont {K.}~\bibnamefont {Luksch}}, \bibinfo
  {author} {\bibfnamefont {S.}~\bibnamefont {Sunami}}, \bibinfo {author}
  {\bibfnamefont {T.~L.}\ \bibnamefont {Harte}}, \bibinfo {author}
  {\bibfnamefont {B.}~\bibnamefont {Yuen}}, \bibinfo {author} {\bibfnamefont
  {C.~J.}\ \bibnamefont {Foot}}, \bibinfo {author} {\bibfnamefont {D.~J.}\
  \bibnamefont {Owens}},\ and\ \bibinfo {author} {\bibfnamefont {J.~M.}\
  \bibnamefont {Hutson}},\ }\bibfield  {title} {\bibinfo {title} {Inelastic
  collisions in radiofrequency-dressed mixtures of ultracold atoms},\ }\href
  {https://doi.org/10.1103/physrevresearch.2.033163} {\bibfield  {journal}
  {\bibinfo  {journal} {Phys. Rev. Research}\ }\textbf {\bibinfo {volume}
  {2}},\ \bibinfo {pages} {033163} (\bibinfo {year} {2020})}\BibitemShut
  {NoStop}%
\bibitem [{\citenamefont {Wang}\ \emph {et~al.}(2025)\citenamefont {Wang},
  \citenamefont {Liu}, \citenamefont {Wang}, \citenamefont {Huang},
  \citenamefont {Chen}, \citenamefont {Zhang},\ and\ \citenamefont
  {Hu}}]{Wang2025}%
  \BibitemOpen
  \bibfield  {author} {\bibinfo {author} {\bibfnamefont {T.}~\bibnamefont
  {Wang}}, \bibinfo {author} {\bibfnamefont {Y.}~\bibnamefont {Liu}}, \bibinfo
  {author} {\bibfnamefont {J.}~\bibnamefont {Wang}}, \bibinfo {author}
  {\bibfnamefont {Y.}~\bibnamefont {Huang}}, \bibinfo {author} {\bibfnamefont
  {W.}~\bibnamefont {Chen}}, \bibinfo {author} {\bibfnamefont {Z.}~\bibnamefont
  {Zhang}},\ and\ \bibinfo {author} {\bibfnamefont {J.}~\bibnamefont {Hu}},\
  }\bibfield  {title} {\bibinfo {title} {Observation of modulation-induced
  {Feshbach} resonance},\ }\href {https://doi.org/10.1103/qrz6-gjp6} {\bibfield
   {journal} {\bibinfo  {journal} {Phys. Rev. Lett.}\ }\textbf {\bibinfo
  {volume} {135}},\ \bibinfo {pages} {203201} (\bibinfo {year}
  {2025})}\BibitemShut {NoStop}%
\bibitem [{\citenamefont {Journeaux}\ \emph {et~al.}(2026)\citenamefont
  {Journeaux}, \citenamefont {Veschambre}, \citenamefont {Lecomte},
  \citenamefont {Uzan}, \citenamefont {Dalibard}, \citenamefont {Werner},
  \citenamefont {Petrov},\ and\ \citenamefont {Lopes}}]{Journeaux2026}%
  \BibitemOpen
  \bibfield  {author} {\bibinfo {author} {\bibfnamefont {A.}~\bibnamefont
  {Journeaux}}, \bibinfo {author} {\bibfnamefont {J.}~\bibnamefont
  {Veschambre}}, \bibinfo {author} {\bibfnamefont {M.}~\bibnamefont {Lecomte}},
  \bibinfo {author} {\bibfnamefont {E.}~\bibnamefont {Uzan}}, \bibinfo {author}
  {\bibfnamefont {J.}~\bibnamefont {Dalibard}}, \bibinfo {author}
  {\bibfnamefont {F.}~\bibnamefont {Werner}}, \bibinfo {author} {\bibfnamefont
  {D.~S.}\ \bibnamefont {Petrov}},\ and\ \bibinfo {author} {\bibfnamefont
  {R.}~\bibnamefont {Lopes}},\ }\bibfield  {title} {\bibinfo {title} {Two-body
  contact dynamics in a {Bose} gas near a {Fano-Feshbach} resonance},\ }\href
  {https://doi.org/10.1103/prf8-3q27} {\bibfield  {journal} {\bibinfo
  {journal} {Phys. Rev. Lett.}\ }\textbf {\bibinfo {volume} {136}},\ \bibinfo
  {pages} {083404} (\bibinfo {year} {2026})}\BibitemShut {NoStop}%
\bibitem [{\citenamefont {Sykes}\ \emph {et~al.}(2017)\citenamefont {Sykes},
  \citenamefont {Landa},\ and\ \citenamefont {Petrov}}]{Sykes2017}%
  \BibitemOpen
  \bibfield  {author} {\bibinfo {author} {\bibfnamefont {A.~G.}\ \bibnamefont
  {Sykes}}, \bibinfo {author} {\bibfnamefont {H.}~\bibnamefont {Landa}},\ and\
  \bibinfo {author} {\bibfnamefont {D.~S.}\ \bibnamefont {Petrov}},\ }\bibfield
   {title} {\bibinfo {title} {Two- and three-body problem with {Floquet-driven}
  zero-range interactions},\ }\href
  {https://doi.org/10.1103/physreva.95.062705} {\bibfield  {journal} {\bibinfo
  {journal} {Phys. Rev. A}\ }\textbf {\bibinfo {volume} {95}},\ \bibinfo
  {pages} {062705} (\bibinfo {year} {2017})}\BibitemShut {NoStop}%
\bibitem [{\citenamefont {Dauer}\ \emph {et~al.}(2025)\citenamefont {Dauer},
  \citenamefont {Pelster},\ and\ \citenamefont {Eggert}}]{Dauer2025}%
  \BibitemOpen
  \bibfield  {author} {\bibinfo {author} {\bibfnamefont {C.}~\bibnamefont
  {Dauer}}, \bibinfo {author} {\bibfnamefont {A.}~\bibnamefont {Pelster}},\
  and\ \bibinfo {author} {\bibfnamefont {S.}~\bibnamefont {Eggert}},\
  }\bibfield  {title} {\bibinfo {title} {Understanding {Floquet} resonances in
  ultracold quantum gas scattering},\ }\href
  {https://doi.org/10.1103/47xs-223h} {\bibfield  {journal} {\bibinfo
  {journal} {Phys. Rev. Lett.}\ }\textbf {\bibinfo {volume} {135}},\ \bibinfo
  {pages} {033402} (\bibinfo {year} {2025})}\BibitemShut {NoStop}%
\bibitem [{\citenamefont {Kaufman}\ \emph {et~al.}(2009)\citenamefont
  {Kaufman}, \citenamefont {Anderson}, \citenamefont {Hanna}, \citenamefont
  {Tiesinga}, \citenamefont {Julienne},\ and\ \citenamefont
  {Hall}}]{Kaufman2009}%
  \BibitemOpen
  \bibfield  {author} {\bibinfo {author} {\bibfnamefont {A.~M.}\ \bibnamefont
  {Kaufman}}, \bibinfo {author} {\bibfnamefont {R.~P.}\ \bibnamefont
  {Anderson}}, \bibinfo {author} {\bibfnamefont {T.~M.}\ \bibnamefont {Hanna}},
  \bibinfo {author} {\bibfnamefont {E.}~\bibnamefont {Tiesinga}}, \bibinfo
  {author} {\bibfnamefont {P.~S.}\ \bibnamefont {Julienne}},\ and\ \bibinfo
  {author} {\bibfnamefont {D.~S.}\ \bibnamefont {Hall}},\ }\bibfield  {title}
  {\bibinfo {title} {Radio-frequency dressing of multiple {Feshbach}
  resonances},\ }\href {https://doi.org/10.1103/physreva.80.050701} {\bibfield
  {journal} {\bibinfo  {journal} {Phys. Rev. A}\ }\textbf {\bibinfo {volume}
  {80}},\ \bibinfo {pages} {050701} (\bibinfo {year} {2009})}\BibitemShut
  {NoStop}%
\bibitem [{\citenamefont {Sawyer}\ \emph {et~al.}(2017)\citenamefont {Sawyer},
  \citenamefont {Horvath}, \citenamefont {Tiesinga}, \citenamefont {Deb},\ and\
  \citenamefont {Kjærgaard}}]{Sawyer2017}%
  \BibitemOpen
  \bibfield  {author} {\bibinfo {author} {\bibfnamefont {B.~J.}\ \bibnamefont
  {Sawyer}}, \bibinfo {author} {\bibfnamefont {M.~S.~J.}\ \bibnamefont
  {Horvath}}, \bibinfo {author} {\bibfnamefont {E.}~\bibnamefont {Tiesinga}},
  \bibinfo {author} {\bibfnamefont {A.~B.}\ \bibnamefont {Deb}},\ and\ \bibinfo
  {author} {\bibfnamefont {N.}~\bibnamefont {Kjærgaard}},\ }\bibfield  {title}
  {\bibinfo {title} {Dispersive optical detection of magnetic {Feshbach}
  resonances in ultracold gases},\ }\href
  {https://doi.org/10.1103/physreva.96.022705} {\bibfield  {journal} {\bibinfo
  {journal} {Phys. Rev. A}\ }\textbf {\bibinfo {volume} {96}},\ \bibinfo
  {pages} {022705} (\bibinfo {year} {2017})}\BibitemShut {NoStop}%
\bibitem [{\citenamefont {Guthmann}\ \emph {et~al.}(2025)\citenamefont
  {Guthmann}, \citenamefont {Lang}, \citenamefont {Kienesberger}, \citenamefont
  {Barbosa},\ and\ \citenamefont {Widera}}]{Guthmann2025}%
  \BibitemOpen
  \bibfield  {author} {\bibinfo {author} {\bibfnamefont {A.}~\bibnamefont
  {Guthmann}}, \bibinfo {author} {\bibfnamefont {F.}~\bibnamefont {Lang}},
  \bibinfo {author} {\bibfnamefont {L.~M.}\ \bibnamefont {Kienesberger}},
  \bibinfo {author} {\bibfnamefont {S.}~\bibnamefont {Barbosa}},\ and\ \bibinfo
  {author} {\bibfnamefont {A.}~\bibnamefont {Widera}},\ }\bibfield  {title}
  {\bibinfo {title} {Floquet engineering of {Feshbach} resonances in ultracold
  gases},\ }\href {https://doi.org/10.1126/sciadv.adw3856} {\bibfield
  {journal} {\bibinfo  {journal} {Sci. Adv.}\ }\textbf {\bibinfo {volume}
  {11}},\ \bibinfo {pages} {eadw3856} (\bibinfo {year} {2025})}\BibitemShut
  {NoStop}%
\bibitem [{\citenamefont {Guthmann}\ \emph {et~al.}(2026)\citenamefont
  {Guthmann}, \citenamefont {Kienesberger}, \citenamefont {Lang}, \citenamefont
  {Lippi},\ and\ \citenamefont {Widera}}]{Guthmann2026}%
  \BibitemOpen
  \bibfield  {author} {\bibinfo {author} {\bibfnamefont {A.}~\bibnamefont
  {Guthmann}}, \bibinfo {author} {\bibfnamefont {L.~M.}\ \bibnamefont
  {Kienesberger}}, \bibinfo {author} {\bibfnamefont {F.}~\bibnamefont {Lang}},
  \bibinfo {author} {\bibfnamefont {E.}~\bibnamefont {Lippi}},\ and\ \bibinfo
  {author} {\bibfnamefont {A.}~\bibnamefont {Widera}},\ }\bibfield  {title}
  {\bibinfo {title} {Engineering a bound state in the continuum via quantum
  interference},\ }\href@noop {} {\  (\bibinfo {year} {2026})},\ \Eprint
  {https://arxiv.org/abs/2602.07809} {arXiv:2602.07809 [cond-mat.quant-gas]}
  \BibitemShut {NoStop}%
\bibitem [{\citenamefont {Autler}\ and\ \citenamefont
  {Townes}(1955)}]{Autler1955}%
  \BibitemOpen
  \bibfield  {author} {\bibinfo {author} {\bibfnamefont {S.~H.}\ \bibnamefont
  {Autler}}\ and\ \bibinfo {author} {\bibfnamefont {C.~H.}\ \bibnamefont
  {Townes}},\ }\bibfield  {title} {\bibinfo {title} {Stark effect in rapidly
  varying fields},\ }\href {https://doi.org/10.1103/physrev.100.703} {\bibfield
   {journal} {\bibinfo  {journal} {Phys. Rev.}\ }\textbf {\bibinfo {volume}
  {100}},\ \bibinfo {pages} {703} (\bibinfo {year} {1955})}\BibitemShut
  {NoStop}%
\bibitem [{\citenamefont {Cohen-Tannoudji}(1996)}]{CohenTannoudji1996}%
  \BibitemOpen
  \bibfield  {author} {\bibinfo {author} {\bibfnamefont {C.~N.}\ \bibnamefont
  {Cohen-Tannoudji}},\ }\bibinfo {title} {{The Autler-Townes Effect
  Revisited}},\ in\ \href {https://doi.org/10.1007/978-1-4612-2378-8_11} {\emph
  {\bibinfo {booktitle} {Amazing Light}}}\ (\bibinfo  {publisher} {Springer New
  York},\ \bibinfo {year} {1996})\ pp.\ \bibinfo {pages} {109--123}\BibitemShut
  {NoStop}%
\bibitem [{\citenamefont {Marte}\ \emph {et~al.}(2002)\citenamefont {Marte},
  \citenamefont {Volz}, \citenamefont {Schuster}, \citenamefont {D\"urr},
  \citenamefont {Rempe}, \citenamefont {van Kempen},\ and\ \citenamefont
  {Verhaar}}]{PhysRevLett.89.283202}%
  \BibitemOpen
  \bibfield  {author} {\bibinfo {author} {\bibfnamefont {A.}~\bibnamefont
  {Marte}}, \bibinfo {author} {\bibfnamefont {T.}~\bibnamefont {Volz}},
  \bibinfo {author} {\bibfnamefont {J.}~\bibnamefont {Schuster}}, \bibinfo
  {author} {\bibfnamefont {S.}~\bibnamefont {D\"urr}}, \bibinfo {author}
  {\bibfnamefont {G.}~\bibnamefont {Rempe}}, \bibinfo {author} {\bibfnamefont
  {E.~G.~M.}\ \bibnamefont {van Kempen}},\ and\ \bibinfo {author}
  {\bibfnamefont {B.~J.}\ \bibnamefont {Verhaar}},\ }\bibfield  {title}
  {\bibinfo {title} {Feshbach resonances in rubidium 87: {P}recision
  measurement and analysis},\ }\href
  {https://doi.org/10.1103/PhysRevLett.89.283202} {\bibfield  {journal}
  {\bibinfo  {journal} {Phys. Rev. Lett.}\ }\textbf {\bibinfo {volume} {89}},\
  \bibinfo {pages} {283202} (\bibinfo {year} {2002})}\BibitemShut {NoStop}%
\bibitem [{Note1()}]{Note1}%
  \BibitemOpen
  \bibinfo {note} {Ref.~\protect \citenum {Ding2017} specifically attributes
  the Feshbach resonance to a bound state in the channel
  $(F_1m_{F_1},F_2m_{F_2})=(20,22)$, where, in reality, the bound state also
  has components from $(11,21)$, $(10,22)$, and $(21,21)$}\BibitemShut
  {NoStop}%
\bibitem [{\citenamefont {Hutson}(2007)}]{Hutson2007}%
  \BibitemOpen
  \bibfield  {author} {\bibinfo {author} {\bibfnamefont {J.~M.}\ \bibnamefont
  {Hutson}},\ }\bibfield  {title} {\bibinfo {title} {Feshbach resonances in
  ultracold atomic and molecular collisions: threshold behaviour and
  suppression of poles in scattering lengths},\ }\href
  {https://doi.org/10.1088/1367-2630/9/5/152} {\bibfield  {journal} {\bibinfo
  {journal} {New J. Phys.}\ }\textbf {\bibinfo {volume} {9}},\ \bibinfo {pages}
  {152} (\bibinfo {year} {2007})}\BibitemShut {NoStop}%
\bibitem [{\citenamefont {Thomas}\ and\ \citenamefont
  {Kj{\ae}rgaard}(2020)}]{Thomas2020}%
  \BibitemOpen
  \bibfield  {author} {\bibinfo {author} {\bibfnamefont {R.}~\bibnamefont
  {Thomas}}\ and\ \bibinfo {author} {\bibfnamefont {N.}~\bibnamefont
  {Kj{\ae}rgaard}},\ }\bibfield  {title} {\bibinfo {title} {{A digital feedback
  controller for stabilizing large electric currents to the ppm level for
  Feshbach resonance studies}},\ }\href {https://doi.org/10.1063/1.5128935}
  {\bibfield  {journal} {\bibinfo  {journal} {Rev. Sci. Instrum.}\ }\textbf
  {\bibinfo {volume} {91}},\ \bibinfo {pages} {034705} (\bibinfo {year}
  {2020})}\BibitemShut {NoStop}%
\bibitem [{\citenamefont {Chilcott}\ \emph {et~al.}(2022)\citenamefont
  {Chilcott}, \citenamefont {Croft}, \citenamefont {Thomas},\ and\
  \citenamefont {Kj\ae{}rgaard}}]{PhysRevA.106.023303}%
  \BibitemOpen
  \bibfield  {author} {\bibinfo {author} {\bibfnamefont {M.}~\bibnamefont
  {Chilcott}}, \bibinfo {author} {\bibfnamefont {J.~F.~E.}\ \bibnamefont
  {Croft}}, \bibinfo {author} {\bibfnamefont {R.}~\bibnamefont {Thomas}},\ and\
  \bibinfo {author} {\bibfnamefont {N.}~\bibnamefont {Kj\ae{}rgaard}},\
  }\bibfield  {title} {\bibinfo {title} {{Microscopy of an ultranarrow Feshbach
  resonance using a laser-based atom collider: A quantum defect theory
  analysis}},\ }\href {https://doi.org/10.1103/PhysRevA.106.023303} {\bibfield
  {journal} {\bibinfo  {journal} {Phys. Rev. A}\ }\textbf {\bibinfo {volume}
  {106}},\ \bibinfo {pages} {023303} (\bibinfo {year} {2022})}\BibitemShut
  {NoStop}%
\bibitem [{\citenamefont {Flash}\ and\ \citenamefont
  {Hogan}(1985)}]{Flash1985}%
  \BibitemOpen
  \bibfield  {author} {\bibinfo {author} {\bibfnamefont {T.}~\bibnamefont
  {Flash}}\ and\ \bibinfo {author} {\bibfnamefont {N.}~\bibnamefont {Hogan}},\
  }\bibfield  {title} {\bibinfo {title} {The coordination of arm movements: an
  experimentally confirmed mathematical model},\ }\href
  {https://doi.org/10.1523/jneurosci.05-07-01688.1985} {\bibfield  {journal}
  {\bibinfo  {journal} {J. Neurosci.}\ }\textbf {\bibinfo {volume} {5}},\
  \bibinfo {pages} {1688} (\bibinfo {year} {1985})}\BibitemShut {NoStop}%
\bibitem [{\citenamefont {Strauss}\ \emph {et~al.}(2010)\citenamefont
  {Strauss}, \citenamefont {Takekoshi}, \citenamefont {Lang}, \citenamefont
  {Winkler}, \citenamefont {Grimm}, \citenamefont {Hecker~Denschlag},\ and\
  \citenamefont {Tiemann}}]{PhysRevA.82.052514}%
  \BibitemOpen
  \bibfield  {author} {\bibinfo {author} {\bibfnamefont {C.}~\bibnamefont
  {Strauss}}, \bibinfo {author} {\bibfnamefont {T.}~\bibnamefont {Takekoshi}},
  \bibinfo {author} {\bibfnamefont {F.}~\bibnamefont {Lang}}, \bibinfo {author}
  {\bibfnamefont {K.}~\bibnamefont {Winkler}}, \bibinfo {author} {\bibfnamefont
  {R.}~\bibnamefont {Grimm}}, \bibinfo {author} {\bibfnamefont
  {J.}~\bibnamefont {Hecker~Denschlag}},\ and\ \bibinfo {author} {\bibfnamefont
  {E.}~\bibnamefont {Tiemann}},\ }\bibfield  {title} {\bibinfo {title}
  {Hyperfine, rotational, and vibrational structure of the
  ${a}^{3}{\ensuremath{\Sigma}}_{u}^{+}$ state of $^{87}\mathrm{Rb}$${}_{2}$},\
  }\href {https://doi.org/10.1103/PhysRevA.82.052514} {\bibfield  {journal}
  {\bibinfo  {journal} {Phys. Rev. A}\ }\textbf {\bibinfo {volume} {82}},\
  \bibinfo {pages} {052514} (\bibinfo {year} {2010})}\BibitemShut {NoStop}%
\bibitem [{Note2()}]{Note2}%
  \BibitemOpen
  \bibinfo {note} {Ref.~\protect \citenum {Ding2017} states a transition
  frequency $\nu =745$~ MHz, slightly lower than what we find, which we
  attribute to using different interaction potentials for the close-coupled
  calculations. We also note that it designates the molecular state as
  $(2-1,22)$, where, in fact, components from
  $(1-1,22)$,$(10,11)$,$(11,20)$,$(21,20)$, and $(11,10)$ will also
  contribute.}\BibitemShut {Stop}%
\bibitem [{\citenamefont {Vanier}\ and\ \citenamefont
  {Audoin}(1989)}]{Vanier1989}%
  \BibitemOpen
  \bibfield  {author} {\bibinfo {author} {\bibfnamefont {J.}~\bibnamefont
  {Vanier}}\ and\ \bibinfo {author} {\bibfnamefont {C.}~\bibnamefont
  {Audoin}},\ }\href {https://doi.org/10.1201/9781003041085} {\emph {\bibinfo
  {title} {{The Quantum Physics of Atomic Frequency Standards}}}}\ (\bibinfo
  {publisher} {CRC Press},\ \bibinfo {year} {1989})\BibitemShut {NoStop}%
\bibitem [{\citenamefont {Cosco}(2023)}]{Cosco2023}%
  \BibitemOpen
  \bibfield  {author} {\bibinfo {author} {\bibfnamefont {A.}~\bibnamefont
  {Cosco}},\ }\emph {\bibinfo {title} {{Optimized production and probing of
  resonantly interacting lithium-chromium Fermi mixtures}}},\ \href@noop {}
  {Master's thesis},\ \bibinfo  {school} {University of Florence} (\bibinfo
  {year} {2023})\BibitemShut {NoStop}%
\bibitem [{\citenamefont {Scazza}\ \emph {et~al.}(2025)\citenamefont {Scazza},
  \citenamefont {Del~Pace}, \citenamefont {Pieri}, \citenamefont {Concas},
  \citenamefont {Kwon},\ and\ \citenamefont {Roati}}]{Scazza2025}%
  \BibitemOpen
  \bibfield  {author} {\bibinfo {author} {\bibfnamefont {F.}~\bibnamefont
  {Scazza}}, \bibinfo {author} {\bibfnamefont {G.}~\bibnamefont {Del~Pace}},
  \bibinfo {author} {\bibfnamefont {L.}~\bibnamefont {Pieri}}, \bibinfo
  {author} {\bibfnamefont {R.}~\bibnamefont {Concas}}, \bibinfo {author}
  {\bibfnamefont {W.~J.}\ \bibnamefont {Kwon}},\ and\ \bibinfo {author}
  {\bibfnamefont {G.}~\bibnamefont {Roati}},\ }\bibfield  {title} {\bibinfo
  {title} {A low-impedance radio-frequency circuit for fast spin manipulations
  in cold alkali atoms},\ }\href {https://doi.org/10.1063/5.0276530} {\bibfield
   {journal} {\bibinfo  {journal} {Rev. Sci. Instrum.}\ }\textbf {\bibinfo
  {volume} {96}},\ \bibinfo {pages} {104713} (\bibinfo {year}
  {2025})}\BibitemShut {NoStop}%
\bibitem [{\citenamefont {Jain}\ \emph {et~al.}(2026)\citenamefont {Jain},
  \citenamefont {Zhang}, \citenamefont {Culemann},\ and\ \citenamefont
  {Preiss}}]{Jain2026}%
  \BibitemOpen
  \bibfield  {author} {\bibinfo {author} {\bibfnamefont {N.}~\bibnamefont
  {Jain}}, \bibinfo {author} {\bibfnamefont {J.}~\bibnamefont {Zhang}},
  \bibinfo {author} {\bibfnamefont {M.}~\bibnamefont {Culemann}},\ and\
  \bibinfo {author} {\bibfnamefont {P.~M.}\ \bibnamefont {Preiss}},\ }\bibfield
   {title} {\bibinfo {title} {Programmable assembly of ground state fermionic
  tweezer arrays},\ }\href {https://doi.org/10.1103/fsmh-dz71} {\bibfield
  {journal} {\bibinfo  {journal} {Phys. Rev. Lett.}\ }\textbf {\bibinfo
  {volume} {136}},\ \bibinfo {pages} {223402} (\bibinfo {year}
  {2026})}\BibitemShut {NoStop}%
\bibitem [{Note3()}]{Note3}%
  \BibitemOpen
  \bibinfo {note} {For a field $B=1007.3$~G, the $|1,1\rangle +|1,0\rangle $
  threshold is located $\sim 605$~MHz above $|1,1\rangle +|1,1\rangle $, making
  the transition energetically allowed for a 751~MHz photon.}\BibitemShut
  {Stop}%
\bibitem [{\citenamefont {Bennett}\ \emph {et~al.}(2017)\citenamefont
  {Bennett}, \citenamefont {Gibble}, \citenamefont {Kokkelmans},\ and\
  \citenamefont {Hutson}}]{Bennett2017}%
  \BibitemOpen
  \bibfield  {author} {\bibinfo {author} {\bibfnamefont {A.}~\bibnamefont
  {Bennett}}, \bibinfo {author} {\bibfnamefont {K.}~\bibnamefont {Gibble}},
  \bibinfo {author} {\bibfnamefont {S.}~\bibnamefont {Kokkelmans}},\ and\
  \bibinfo {author} {\bibfnamefont {J.~M.}\ \bibnamefont {Hutson}},\ }\bibfield
   {title} {\bibinfo {title} {Atomic clock measurements of quantum scattering
  phase shifts spanning {Feshbach} resonances at ultralow fields},\ }\href
  {https://doi.org/10.1103/physrevlett.119.113401} {\bibfield  {journal}
  {\bibinfo  {journal} {Phys. Rev. Lett.}\ }\textbf {\bibinfo {volume} {119}},\
  \bibinfo {pages} {113401} (\bibinfo {year} {2017})}\BibitemShut {NoStop}%
\bibitem [{Note4()}]{Note4}%
  \BibitemOpen
  \bibinfo {note} {For a $B_{\protect \rm rf}=2.2$~G amplitude both our
  calculations and our measured tuning rate $\alpha $ predict a splitting of
  $\sim 0.5$~G. This mutual agreement does not extend to the prediction of
  Ref.~\protect \rev@citealp {Ding2017}, where a $\sim 0.5$~G splitting is
  obtained for $B_{\protect \rm rf}=5$~G.}\BibitemShut {Stop}%
\bibitem [{\citenamefont {Maury}\ \emph {et~al.}(2023)\citenamefont {Maury},
  \citenamefont {Bakkali-Hassani}, \citenamefont {Chauveau}, \citenamefont
  {Rabec}, \citenamefont {Nascimbene}, \citenamefont {Dalibard},\ and\
  \citenamefont {Beugnon}}]{Maury2023}%
  \BibitemOpen
  \bibfield  {author} {\bibinfo {author} {\bibfnamefont {C.}~\bibnamefont
  {Maury}}, \bibinfo {author} {\bibfnamefont {B.}~\bibnamefont
  {Bakkali-Hassani}}, \bibinfo {author} {\bibfnamefont {G.}~\bibnamefont
  {Chauveau}}, \bibinfo {author} {\bibfnamefont {F.}~\bibnamefont {Rabec}},
  \bibinfo {author} {\bibfnamefont {S.}~\bibnamefont {Nascimbene}}, \bibinfo
  {author} {\bibfnamefont {J.}~\bibnamefont {Dalibard}},\ and\ \bibinfo
  {author} {\bibfnamefont {J.}~\bibnamefont {Beugnon}},\ }\bibfield  {title}
  {\bibinfo {title} {Precision measurement of atom-dimer interaction in a
  uniform planar bose gas},\ }\href
  {https://doi.org/10.1103/physrevresearch.5.l012020} {\bibfield  {journal}
  {\bibinfo  {journal} {Phys. Rev. Research}\ }\textbf {\bibinfo {volume}
  {5}},\ \bibinfo {pages} {$L012020$} (\bibinfo {year} {2023})}\BibitemShut
  {NoStop}%
\bibitem [{\citenamefont {Ballu}\ \emph {et~al.}(2025)\citenamefont {Ballu},
  \citenamefont {Yao}, \citenamefont {Mirmand}, \citenamefont {Papoular},
  \citenamefont {Perrin},\ and\ \citenamefont {Perrin}}]{Ballu2025}%
  \BibitemOpen
  \bibfield  {author} {\bibinfo {author} {\bibfnamefont {M.}~\bibnamefont
  {Ballu}}, \bibinfo {author} {\bibfnamefont {Z.}~\bibnamefont {Yao}}, \bibinfo
  {author} {\bibfnamefont {B.}~\bibnamefont {Mirmand}}, \bibinfo {author}
  {\bibfnamefont {D.~J.}\ \bibnamefont {Papoular}}, \bibinfo {author}
  {\bibfnamefont {H.}~\bibnamefont {Perrin}},\ and\ \bibinfo {author}
  {\bibfnamefont {A.}~\bibnamefont {Perrin}},\ }\bibfield  {title} {\bibinfo
  {title} {Microwave spectroscopy of ultracold-sodium least-bound molecular
  states},\ }\href {https://doi.org/10.1103/qv5s-71ll} {\bibfield  {journal}
  {\bibinfo  {journal} {Phys. Rev. A}\ }\textbf {\bibinfo {volume} {112}},\
  \bibinfo {pages} {013312} (\bibinfo {year} {2025})}\BibitemShut {NoStop}%
\bibitem [{\citenamefont {Sawyer}\ \emph {et~al.}(2019)\citenamefont {Sawyer},
  \citenamefont {Chilcott}, \citenamefont {Thomas}, \citenamefont {Deb},\ and\
  \citenamefont {Kjærgaard}}]{Sawyer2019}%
  \BibitemOpen
  \bibfield  {author} {\bibinfo {author} {\bibfnamefont {B.~J.}\ \bibnamefont
  {Sawyer}}, \bibinfo {author} {\bibfnamefont {M.}~\bibnamefont {Chilcott}},
  \bibinfo {author} {\bibfnamefont {R.}~\bibnamefont {Thomas}}, \bibinfo
  {author} {\bibfnamefont {A.~B.}\ \bibnamefont {Deb}},\ and\ \bibinfo {author}
  {\bibfnamefont {N.}~\bibnamefont {Kjærgaard}},\ }\bibfield  {title}
  {\bibinfo {title} {Deterministic quantum state transfer of atoms in a random
  magnetic field},\ }\href {https://doi.org/10.1140/epjd/e2019-90541-2}
  {\bibfield  {journal} {\bibinfo  {journal} {Eur. Phys. J. D}\ }\textbf
  {\bibinfo {volume} {73}},\ \bibinfo {pages} {160} (\bibinfo {year}
  {2019})}\BibitemShut {NoStop}%
\bibitem [{\citenamefont {Thomas}(2020)}]{thomas2020coupled}%
  \BibitemOpen
  \bibfield  {author} {\bibinfo {author} {\bibfnamefont {R.~J.}\ \bibnamefont
  {Thomas}},\ }\href@noop {} {\bibinfo {title} {Coupled channels solver}},\
  \bibinfo {howpublished}
  {\url{https://github.com/kjaergaard-lab/coupled-channels-solver}} (\bibinfo
  {year} {2020})\BibitemShut {NoStop}%
\end{thebibliography}%

\newpage
\clearpage

\makeatletter
\normalsize
\onecolumngrid

\begin{figure}[h]
\centering \includegraphics[width=0.9\columnwidth]{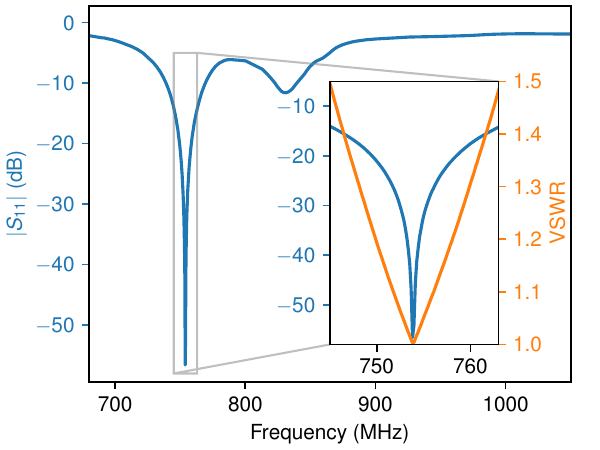}
\caption{Measured reflection coefficient $|S_{11}|$ of the rf antenna as a function of frequency. Inset: $|S_{11}|$ and VSWR over the 745--763 MHz region.} 
\label{fig:antenna_s11}
\end{figure}

\begin{center}{{\bf End Matter}}\end{center}
\twocolumngrid
\textit{Atomic sample preparation---} We load an evaporatively cooled sample of $^{87}\mathrm{Rb}$ atoms in the
$\lvert F=2,m_F=2\rangle$ hyperfine state from a magnetic trap into a crossed optical dipole trap. The crossed trap is formed by two
 $1064~\mathrm{nm}$ laser beams
with a waist of \SI{\sim 60}{\micro\metre}, intersecting orthogonally at the center of a rectangular cross section ($12.4\times22.4$~mm) glass cell (see \fref{fig:setup}). Atoms are transferred to the $\lvert F=1,m_F=1\rangle$ state
 by microwave adiabatic rapid passage \cite{Sawyer2019}. 
\newline\\
\textit{Bound states---}\Fref{fig:bound_states} shows the results of calculating $\rm^{87}Rb_2$ bound state energies (relative to the $\ket{1,1}+\ket{1,1}$ threshold) as a function of magnetic field. States that are not relevant  to the two-level framework are shown as faded  \cite{thomas2020coupled}.
\\\\\\\
\textit{Circuit board antenna---}Using a vector network analyzer we measure the input reflection coefficient $S_{11}$ for the PCB antenna used for rf-dressing (see \fref{fig:antenna_s11}). The operating bandwidth as defined by a VSWR of less than 1.5 is 745--763~MHz.

\textit{Magnetic amplitude of rf field---}\begin{figure}[tb!]
    \centering
    \includegraphics[width=0.9\linewidth]{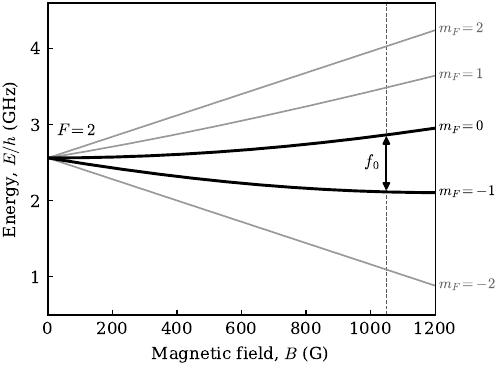}
\caption{Breit--Rabi diagram of the $^{87}$Rb $F=2$ ground-state manifold.
States are labeled by the low-field $m_F$ to which they connect adiabatically.
The arrow marks the rf transition between the states connected to
$\ket{2,0}$ and $\ket{2,-1}$, driven at $f_0 = 749.715$~MHz to calibrate the
rf field amplitude. The dashed line indicates the bias field
$B = 1048.83$~G used for the calibration.}
\label{fig:Rabi_measurements2}
\end{figure}To estimate the rf magnetic field amplitude at the position of the atoms when the loop antenna is driven at $\nu_{\rm rf}\sim750$~MHz, we addressed the atomic hyperfine transition between the states adiabatically connected to the $\lvert F=2,m_F=0\rangle$ and $\lvert F=2,m_F=-1\rangle$ at a bias field of $B=1048.83~\mathrm{G}$ (see \fref{fig:Rabi_measurements2}). State transfer was detected by applying a magnetic-field gradient and measuring the resulting displacement of the atomic cloud. The two states have different magnetic moments and therefore experience different forces in the gradient; the measured displacement was consequently used as a proxy for the transferred population. We first recorded a Rabi spectrum by varying the rf frequency for a fixed pulse duration of \SI{0.6}{\micro\second} [see \fref{fig:Rabi_measurements}(a)]. A fit to the finite-pulse Rabi line shape gave a resonance frequency of $f_0=749.715(4)~\mathrm{MHz}$.
The rf frequency was subsequently fixed at  $749.715~\mathrm{MHz}$, and the pulse duration varied to measure Rabi oscillations [see \fref{fig:Rabi_measurements}(b)]. These were obtained by driving our loop antenna with our maximally available rf power ($\sim 100$~W). The time variation of the displacement was fitted using $\Delta z(t) \propto \sin^{2}\!\left( \Omega_{\rm HFS}t/2\right)$, where $\Omega_{\rm HFS}$ is the Rabi frequency.

From the fitted Rabi frequency $\Omega_{\mathrm{HFS}}=2\pi\times0.9774(2) \,\mathrm{MHz}$ we can work out the amplitude of the corresponding driving field. We numerically diagonalize the $^{87}\mathrm{Rb}$ ground-state hyperfine-Zeeman Hamiltonian at $B=1048.83$~G and calculate the magnetic dipole matrix element between the two energy eigenstates that adiabatically connect to $\ket{2,-1}$ and $\ket{2,0}$ at zero field. The magnetic rf field amplitude can then be obtained from:
\begin{equation}\label{Eq:1_Brf_magnitude}
B_{\mathrm{rf}}=\hbar\Omega_{\mathrm{HFS}}/\mu_B \kappa,
\end{equation}
where $\mu_B$ is the Bohr magneton and $\kappa=0.6137$ is the calculated, dimensionless matrix element of the transition. We repeat this for the Rabi frequencies fitted in several time windows of the rf pulse (0--3, 10--13, 20--23 and 50--53~$\mu$s) and take the mean of the resulting $B_{\mathrm{rf}}$ values, with their standard deviation as the uncertainty.
The magnetic rf field amplitude is found to be $B_{\mathrm{rf}}=1.14(11)~\mathrm{G}$ at our maximum drive.
\begin{figure}[tb!]
    \centering
    \includegraphics[width=\linewidth]{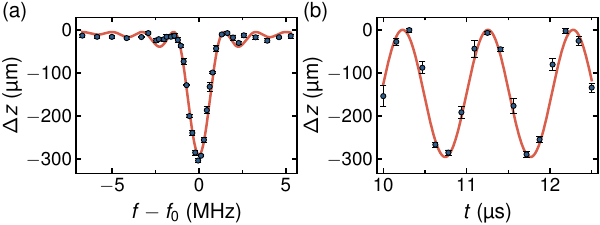}
    \caption{Spectroscopy on the transition marked in \fref{fig:Rabi_measurements2}. (a) Atomic cloud displacement $\Delta z$ as a function of rf detuning $f-f_0$, measured in the presence of a magnetic-field gradient using a \SI{0.6}{\micro\second} rf pulse. The solid curve is a fit to the finite-pulse Rabi line shape ($f_0=749.715(4)~\mathrm{MHz}$). (b) Atomic cloud displacement as a function of rf pulse duration for $f=749.715~\mathrm{MHz}$. The solid curve is a sinusoidal fit to the resulting resonant Rabi oscillations.}
    \label{fig:Rabi_measurements}
\end{figure}
For \fref{fig:rf_loss_B1007G}(a) and (b), the rf magnetic-field amplitude, $B_{\mathrm{rf}}$, at each rf power was determined from the corresponding fitted Rabi frequency using \eref{Eq:1_Brf_magnitude}. Since the variation of $B_{\mathrm{rf}}$ between time windows was measured only at our maximum drive, we used its standard deviation, $0.11~\mathrm{G}$, as an upper bound for the uncertainty of $B_{\mathrm{rf}}$ at all powers.

\textit{Estimation of $\alpha_{\rm calc}$}---The rf couples the $M=2$ closed-channel state $|\psi_{M=2}\rangle$ to an $M=1$ state $|\psi_{M=1}\rangle$ (\fref{fig:bound_states}). Without rf, the resonance lies at magnetic field $B=B_0$, where $|\psi_{M=1}\rangle$ is $h\nu_0$ below $|\psi_{M=2}\rangle$. In a two-level toy model with basis $\{|\psi_{M=2},N\rangle,\allowbreak\,|\psi_{M=1},N{+}1\rangle\}$, where $N$ is the number of rf photons, the Hamiltonian near $B_0$ in the rotating-wave approximation is
\begin{equation}
H=\begin{pmatrix}
E_1 & \hbar\Omega_{\rm mol}/2\\
\hbar\Omega_{\rm mol}/2 & E_2
\end{pmatrix},
\label{Eq:two_level_H}
\end{equation}
where the diagonal energies, relative to the $|1,1\rangle+|1,1\rangle$ threshold, are $E_1=\delta\mu_1(B-B_0)$ and $E_2=\delta\mu_2(B-B_0)+h(\nu_{\rm rf}-\nu_0)$. Here $\delta\mu_1$ and $\delta\mu_2$ are the differential magnetic moments of the $M=2$ and $M=1$ states, respectively, relative to that of a free $|1,1\rangle+|1,1\rangle$ atom pair. The off-diagonal coupling contains $\hbar\Omega_{\rm mol}=\mu_B\kappa_{\rm mol}B_{\rm rf}$, the molecular counterpart of \eref{Eq:1_Brf_magnitude}, where $\kappa_{\rm mol}$ is the dimensionless magnetic-dipole transition matrix element between $|\psi_{M=2}\rangle$ and $|\psi_{M=1}\rangle$. A resonance occurs where a dressed state reaches threshold, i.e., where an eigenvalue of $H$ vanishes: $\det H=E_1E_2-(\hbar\Omega_{\rm mol}/2)^2=0$. At $\nu_{\rm rf}=\nu_0$ the two solutions are separated by $\hbar\Omega_{\rm mol}/\sqrt{\delta\mu_1\delta\mu_2}$. Since $\Omega_{\rm mol}\propto B_{\rm rf}$, the splitting is $\Delta B=\alpha_{\rm calc}B_{\rm rf}$, where
\begin{equation}
\alpha_{\rm calc}=\frac{\mu_B\kappa_{\rm mol}}{\sqrt{\delta\mu_1\delta\mu_2}}=0.205,
\end{equation}
for $\kappa_{\rm mol}=0.531$, $\delta\mu_1/h=3.96$~MHz/G, and $\delta\mu_2/h=3.33$~MHz/G from our coupled-channels solver \cite{thomas2020coupled}. 
\renewcommand{\refname}{\textbf{Supplementary References}}

\end{document}